\documentclass{ar-1col}
\usepackage{url}
\usepackage[numbers,sort&compress]{natbib}
\usepackage{amsmath,amssymb}
\usepackage{bm,bbm}

\usepackage[colorlinks,linkcolor={blue},citecolor={blue}, urlcolor={blue}]{hyperref}

\definecolor{armygreen}{rgb}{0.55, 0.73, 0.0}
\definecolor{darkblue}{rgb}{0.0, 0.0, 0.41}

\usepackage{soul}

\usepackage{xcolor}

\newcommand{\ejd}[1]{#1}
\newcommand{\msd}[1]{#1}

\newcommand{\sonia}[1]{#1}

\jname{Xxxx. Xxx. Xxx. Xxx.}
\jvol{AA}
\jyear{YYYY}
\doi{10.1146/((please add article doi))}

\begin{document}

\markboth{Dresselhaus et al.}{Textiles, Topology, and Mechanics}

\title{Textiles: from twisted yarn to topology and mechanics}


\author{Elizabeth~J.~Dresselhaus,$^{1, 4}$ Sonia~Mahmoudi,$^{2,3,4}$ Lauren~Niu,$^{5,6}$ Samuel~Poincloux,$^7$ Vanessa~Sanchez,$^{8,9,10}$ and~Michael~S.~Dimitriyev$^{11}$
\affil{$^1$Department of Chemical and Biomolecular Engineering, University of California, Berkeley, CA 94720, USA}
\affil{$^2$Advanced Institute for Materials Research (WPI-AIMR), Tohoku University, 2-1-1 Katahira, Aoba-ku, Sendai, Miyagi 980-8577, Japan}
\affil{$^3$RIKEN, Center for Interdisciplinary Theoretical and Mathematical Sciences (iTHEMS), 2-1, Hirosawa, Wako, Saitama 351-0198, Japan}
\affil{$^4$International Institute for Sustainability with Knotted Chiral Meta Matter (WPI-SKCM$^2$), Hiroshima University, 1-3-1 Kagamiyama, Higashi-Hiroshima, Hiroshima 739-8526, Japan}
\affil{$^5$Center for Functional Fabrics, Drexel University, Philadelphia, USA, 19104}
\affil{$^6$Department of Physics and Astronomy, University of Pennsylvania, Philadelphia, USA, 19104}
\affil{$^7$Department of Physical Sciences, Aoyama Gakuin University, 5-10-1 Fuchinobe, Sagamihara, Kanagawa 252-5258, Japan}
\affil{$^8$Department of Mechanical Engineering, Rice University, Houston, Texas, USA, 77005}
\affil{$^9$Department of Materials Science and NanoEngineering, Rice University, Houston, Texas, USA, 77005} 
\affil{$^{10}$Rice Advanced Materials Institute, Rice University, Houston, Texas, USA, 77005} 
\affil{$^{11}$Department of Materials Science and Engineering, Texas A\&M University, College Station, Texas, USA, 77843; email: msdim@tamu.edu}
}

\begin{abstract}
While textiles have existed throughout much of human history as complex mechanical metamaterials, textile science has largely been overlooked by the physics community until recently. In this review, we consider the symmetry, topology, and mechanics of woven and knitted materials, showing that they represent a unique, if under-explored, \msd{corner} of condensed matter. We start with the basic construction and \msd{elementary mechanical model} of spun yarn, reviewing recent developments twisted bundle structures. We then introduce woven and knitted fabrics as materials with layer symmetries that can be topologically characterized as knots and links in the thickened torus.
We finally discuss fabric mechanics and geometry\msd{, invoking recent results surrounding yarn-level and stitch-level geometry, deformations}, and defect structures. 
\end{abstract}

\maketitle

\tableofcontents

\section{INTRODUCTION}

Humans have used textiles in some form for much of our collective history.
From spinning grasses and animal hairs to make stronger threads, yarns, and ropes, to industrial-scale weaving of bespoke patterns with the Jacquard loom, and finally to the technical textiles and smart wearables found today, the methods and materials used for the creation of textiles have undergone a dramatic evolution.
Nonetheless, the themes of twisting, knotting, and entangling as a way of combining fiber-like constituents remains unchanged.
While we will not attempt a historical introduction to textiles, we refer the interested reader to refs.~\cite{postrel2020fabric, finlay2021fabric, stclair2019goldenthread}.

In recent years, physicists have taken an interest in textiles.
Textile science and engineering is a mature field that has developed standard measures, metrology methods, and models that have enabled the industry to grow \msd{and occupy a pivotal role in global trade.}
\msd{Meanwhile,} fiber artists and the crafter community ensure that traditional methods of weaving, knitting, crocheting, and other techniques persist, while finding new ways of building complex patterns and mechanical behaviors into fabrics.
There are at least two reasons for physicists to pick up this thread: one, through the exploration of mechanical metamaterials and inverse design; two, through an interest in mathematical modeling towards developing structure-function relationships across scales.
Indeed, considerable progress has been made in both of these fields over the past few decades~\cite{Panagiotou2013,Panagiotou2015,Panagiotou2024,Fukuda2026,Grunbaum1980,Grunbaum1988,Markande2020,Liu2018,OKeeffe2020,Grishanov2007,Morton2009,Bright2020,Diamantis2024,Grishanov2009I,Grishanov2009II,Grishanov.Vassiliev1,Grishanov.Vassiliev2,Kawauchi2018}.
In this Review, we will consider textiles through the lens of condensed matter physics, focusing on topology, symmetry, and non-linear mechanics.
In doing so, we offer an invitation for others to explore this fascinating topic.

In order for physicists to contribute to the study and tradition of textiles, it is helpful to learn some of the basic language.
Textile terminology has a complex history and even has differences amongst traditional fiber artists and industrial-scale textile engineers, as well as across regions, reflecting the global and culturally interwoven nature of textile practices; a minimal dictionary is provided in the margin.
A classical example of this is the use of ``woven'' to describe a textile whose structure consists of yarn that crosses the fabric in a well-prescribed manner based on a specific pattern.
By contrast, a ``non-woven'' can be defined as any other textile involving some degree of randomness to the microstructure, including felt---a material made by pressing fibers together and held together by inter-fiber friction.
However, while this definition would include knitted and crocheted fabrics as wovens, these have structures distinct from those created through a weaving process. 
Notably, even within the textile field, terminology can remain unsettled. 
For example, the definition of ``nonwoven fabric" in the Textile Institute's Textile Terms and Definitions begins with the phrase ``Opinions vary," reflecting on several open disagreements in the field, including whether ``nonwoven fabric" refers broadly to any fabric that is \textit{not formed through weaving}, including knits, braids, etc., or more specifically to structures formed directly from fibers rather than yarns~\cite{textile_institute_1995_terms}.  
To avoid confusion, we will adopt the terminology of Hearle, Grosberg, and Backer~\cite{hearle1969structural} and call these prescribed assemblies of yarns ``interlaced fabrics.'' 

\begin{marginnote}[0pt]
\entry{Interlaced fabric}{A fabric created from crossing yarns in a specific pattern; often called a ``woven'' but does not necessarily need to be created by weaving.}
\entry{Spinning} {Process of forming a yarn from fibers by applying tension and twist.}
\entry{tex}{Unit of linear mass density; $1\,{\rm tex} = 1\,{\rm g/km}$}
\entry{denier}{Unit of linear mass density; $9\,{\rm den} = 1\,{\rm tex}$}
\end{marginnote}


\section{YARN: A BASIC STRUCTURAL ELEMENT}

Yarn, threads, and strings all describe the basic structural element of \msd{interlaced} fabrics; for brevity, \msd{we will just use the term ``yarn.''}
\msd{Yarn generally refers to the material used directly in fabric creation; different terms are used throughout the yarn creation process}~\cite{textile_institute_1995_terms}.
Traditionally created from fibers consisting of either animal products (e.g.~wool and silk) or plant material (e.g.~cotton, hemp, linen), modern synthetic yarns are formed from fibers formed when a polymer melt (e.g.~polyacrylate, polyester, rayon, nylon) is extruded through a narrow pore. 
Semi-synthetic ``regenerated fibers'' harvest and reconstruct cellulose (e.g.~from bamboo) to create rayon and associated fibers.
Finally, there is increasing interest in the design of shape-actuating fabrics whose yarn is based on shape-changing materials \cite{Sanchez2021} such as shape memory alloys~\cite{weinberg2020multifunctional,Abel2013}, liquid crystal elastomers~\cite{Escobar2026}, as well as yarn constructed from human cells for tissue engineering~\cite{Magnan2020}. 

Since natural fibers tend to be too short (``filament silk'' is an exception) for fabric creation, yarn is \msd{often} a hierarchical bundle of \msd{fibers} that are \textit{spun} together under tension.
In contrast, polymer extrusion can create long \textit{filaments} that extend the full length of the yarn. 
Filament materials, which can range from polymers, including elastomers, to metallic wires, can be used as the yarn itself (``monofilament'') or can be wrapped into a bundle (``multifilament'').

Given the variability of yarn material and structure, it is difficult to arrive at a standard characterization of yarn properties.
For instance, the diameters of fluffy, multifilament bundles are not \msd{well defined}, and they are rather characterized by their linear mass density (see \textit{tex}, \textit{denier} in the textile lexicon).
As such, beyond simple monofilament, \msd{much is unknown} about the mechanical (i.e.~stress-vs-strain) constitutive relations of various yarn types.
Moreover, contact mechanics involves the added complexity of friction and contact geometry.
While experimental characterization (e.g.~tensile and flexural testing) can provide useful details, yarn properties are often sensitive to a history of loading and unloading, degradation, and even humidity in the case of many natural fibers.

\subsection{Hierarchical, plied yarn}

\begin{figure}
\includegraphics[width=\textwidth]{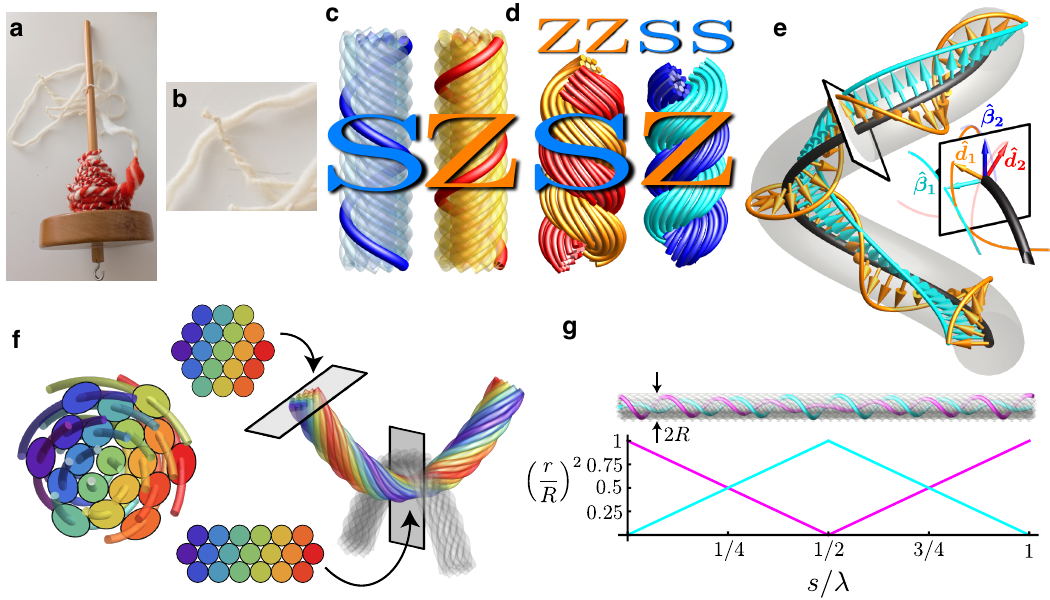}
\caption{Geometry of yarn. (a) Hand-spun yarn on a drop spindle. White and orange single-ply yarns are S-spun separately and then plied together with Z-spin. This yarn was made from Shetland wool staple fibers. (b) Close-up of (a) along the single-ply white yarn reveals the plectoneme instability of unbalanced yarn. (c) S and Z variants of spun yarn, with multi-plied bundles (superhelices) shown in (d). (e) Depiction of the Bishop frame $\left\{\bm{\hat{\beta}_1},\bm{\hat{\beta}_2}\right\}$ and the material frame $\left\{\bm{\hat{d}_1},\bm{\hat{d}_2}\right\}$ adapted to the centerline of a yarn segment. (f) Right: Changes in yarn geometry and fiber packing while two yarn segments are in contact: un-compressed yarn adopts an idealized circular cross-section with hexagonal packing of the constituent fibers, whereas in the contact region, the cross-sections become anisotropic but may still maintain a hexagonal packing. Left: Alterations to the hexagonal packing arrangement arising from twist-induced distortions of the fiber cross sections. (g) Schematic of fiber migration where a pair of filaments transitions between the outer layer and inner core of a spun yarn segment.}
\label{fig:1}
\end{figure}

Yarns and threads made from animal and plant fibers are complex, hierarchical structures. 
Staple fibers are short, often only a few centimeters in length~\cite{Warren2018}.
Consequently, in order to create yarn of arbitrary length, these staple fibers are combined using \msd{through ``spinning''---a process used by humans} for millennia, developed separately amongst many cultures~\cite{Ryder1968}.
While modern industrial methods have automated spinning, the basic process remains the same: fibers are aligned and formed into a bundle; tension is applied across the bundle; the bundle is twisted under tension; more fibers are added to allow the bundle to grow in length.
A drop spindle, such as the one shown in \textbf{figure~\ref{fig:1}\textit{a}}, uses gravity and rotational inertia to apply tension and torque to the fibers. 
By twisting under tension, the \msd{bundle maintains its form through inter-fiber friction}, resulting in a single ``ply'' of yarn (known as singles yarn) as seen in \textbf{figure~\ref{fig:1}\textit{b}}.

\msd{Spinning results in the buildup of residual torque} arising from the imposed helical shape of the fibers; the spinning direction chooses the handedness of the helix---designated ``S'' or ``Z,'' as shown in \textbf{figure~\ref{fig:1}\textit{c}}---and thus the sign of the torque.
Such ``single-ply'' yarns are susceptible to relaxing residual torque through a mechanical instability that converts \msd{the bundle twist to} \textit{writhe} (three-dimensional coiling of the bundle).
This helical \msd{\emph{twist-to-writhe}} instability, also known as the ``plectoneme'' instability, in reference to a similar phenomenon in DNA supercoiling, or the ``phone cord'' instability~\cite{Podvratnik2011,Audoly2010}, can distort the appearance of the resulting textile.
\msd{Similar phenomena can be seen in nylon thread~\cite{Haines2014}, carbon nanotube yarns~\cite{Shang2013}, and nanorods~\cite{Wang2011}.}
\msd{This breaks mirror symmetries in the fabric. This can cause the rest configuration of knitted fabric to adopt a preferred in-plane shear, known as fabric ``spirality''~\cite{Tao1997}}.
In woven fabrics, \msd{opposing corners of the fabric curl up and down, out of the plane of the fabric~\cite{Whitman1947}.}
In order to \msd{avoid or mitigate this instability and add strength to the yarn}, multiple plies are \textit{counter-twisted} together to create a ``multi-ply'' or ``balanced'' yarn, as in \textbf{figure~\ref{fig:1}\textit{d}}.
\msd{Twisted monofilament bundles can also be balanced by counter-twisting multiple bundles together.}


\subsection{Kirchhoff rod model}

\msd{The simplest model of yarn abstracts it as a 1-dimensional curve that can bend but cannot change its length (i.e.~an inextensible elastica model).}
\msd{This does not account for yarn's axial compressibility or any other details arising from spinning or bundled structure.}
\msd{In this and related elastic curve models, the yarn is described as a thickened space curve $\bm{\Gamma}(s)$, where $s \in [0,L]$ is an arclength parameter defined such that $\partial_s \bm{\Gamma} = \hat{\mathbf{t}}(s)$ is the unit tangent vector (see e.g.,~\cite{KamienRMP}).}
\msd{This ``center curve''} is taken to lie along the center of mass of each small segment of the yarn throughout its length~\cite{Audoly2010}.
The bending response of the yarn depends on the curvature $\kappa(s)$ of $\bm{\Gamma}$, given by $\partial_{ss}\bm{\Gamma} = \partial_s\hat{\mathbf{t}} = \kappa\hat{\mathbf{n}}$, where $\hat{\mathbf{n}}(s)$ is the unit normal vector.
The twisting response requires \msd{augmenting the curve with} an orthonormal \textit{material frame} $\left\{\hat{\mathbf{d}}_1,\hat{\mathbf{d}}_2\right\}$ where $\hat{\mathbf{d}}_\mu\cdot \hat{\mathbf{t}} = 0$ at each point $s$.
The instantaneous twist of the yarn $\Omega$ at each point along the curve is given by $\Omega(s) \equiv \hat{\mathbf{d}}_2\cdot\partial_s\hat{\mathbf{d}}_1$.
For more about the definition of the material frame, see the sidebar below.

Kirchhoff rod theory, \msd{which describes space curves with bending and twisting stiffness, might be considered the ``harmonic oscillator'' of yarn mechanics}.
While \msd{ a more general version would incorporate anisotropies in bending and twisting moduli as well as preferred yarn curvature}, for most types of yarn, it is typical that (i) the cross-section is approximately circular, (ii) gently twisting the yarn does not bend it, and (iii) there may be a preferred twist $\overline{\Omega}$ in the case of un-balanced or pre-twisted yarn.
This results in a more typical yarn elastic energy of the form
\begin{equation}
E = \frac{1}{2}\int{\rm d}s\,\left[ B\,\kappa^2(s) + J\left(\Omega(s) - \overline{\Omega}\right)^2\right]\, ,
\end{equation}
where $B$ and $J$ are, respectively, the bending and twisting moduli of the yarn and can be measured for both monofilament and spun yarn~\cite{corn,park_bending_2006,Zurek1980}.
An additional stretching cost can be added to this model, but for most spun yarns, this cost is prohibitively large, and the yarn can be assumed to maintain a fixed length; elastomeric yarns are a notable exception.

\begin{textbox}[h]
\section{THE FRAMING OF A CURVE}
\msd{When defining a frame adapted to a space curve, it is convenient to align one of the frame vectors with the tangent unit vector $\hat{\mathbf{t}}$.}
\msd{The \textit{Frenet-Serret frame}, consists} of the tangent vector $\hat{\mathbf{t}}$, the normal vector $\hat{\mathbf{n}} \equiv \partial_s \hat{\mathbf{t}}/\left|\partial_s \hat{\mathbf{t}}\right|$, and the binormal vector $\hat{\mathbf{b}} \equiv \hat{\mathbf{t}}\times\hat{\mathbf{n}}$.
\msd{The frame's infinitesimal rotation rate is} given by $\partial_s\left\{\hat{\mathbf{t}},\hat{\mathbf{n}},\hat{\mathbf{b}}\right\} = \bm{\omega}_{\rm FS}\times\left\{\hat{\mathbf{t}},\hat{\mathbf{n}},\hat{\mathbf{b}}\right\}$, where the Darboux vector \msd{$\bm{\omega}_{\rm FS}= \tau\hat{\mathbf{t}} + \kappa\hat{\mathbf{b}}$ involves the curvature $\kappa$ and torsion $\tau$; this definition fails in straight regions where $\kappa = 0$.}
\msd{To fix this, the \textit{Bishop frame}} $\left\{\hat{\mathbf{t}},\hat{\bm{\beta}}_1,\hat{\bm{\beta}}_2\right\}$ imposes an additional constraint of zero twist, $\hat{\bm{\beta}}_1\cdot\partial_s\hat{\bm{\beta}}_2=0$, resulting in a Darboux vector $\bm{\omega}_{\rm B} = -\hat{\bm{\beta}}_1\kappa \sin\left( \int^s_0{\rm d}s'\tau\right)  + \hat{\bm{\beta}}_2\kappa \cos \left( \int^s_0{\rm d}s'\tau\right)$~\cite{Bishop1975}.
\msd{Any other frame $\left\{\hat{\mathbf{t}},\hat{\mathbf{d}}_1,\hat{\mathbf{d}}_2\right\}$, including the \textit{material frame}, can be represented via a rotation of the Bishop frame by an angle $\theta$, as shown in \textbf{figure~\ref{fig:1}\textit{e}}.}
\msd{This rotation is an SO(2) degree of freedom, resulting in a gauge transformation of the frame and a transformed Darboux vector $\bm{\omega} = \Omega\hat{\mathbf{t}} + \kappa \sin\Phi\, \hat{\mathbf{d}}_1 + \kappa\cos\Phi\,\hat{\mathbf{d}}_2$ where $\Omega = \partial_s\theta$ is the relative frame twist and $\Phi \equiv \int^s_0{\rm d}s'(\Omega - \tau)$.}
\end{textbox}

\subsection{Fiber packing}\label{sec:fiber_packing}

\msd{It} is reasonable to model the undeformed state of spun yarn as a straight line with zero curvature ($\kappa = 0$) and a constant twist of the material frame ($\Omega \neq 0$).
While under tension, these filaments will be subject to an aligning force that tends to minimize their distance from the yarn's central axis; due to hard-core interactions between the filaments, they cannot all pack along the central axis.
Consequently, \msd{the tension applies} an effective pressure or confining force~\cite{seguin2022twist} on the filaments in the yarn's cross-section.
\msd{This causes filaments to pack in a way that minimizes the area fraction of voids within the yarn's cross-section, resulting in the idealized model of hexagonal packing, as shown in \textbf{figure~\ref{fig:1}\textit{f}}.}

The hexagonal packing model is largely contingent on the assumption that cutting yarn perpendicular to its local tangent vector gives a 2D lattice of circular disks.
However, due to the twist of the individual filaments about the central axis of the yarn, \msd{filament profiles in this cross-section of the yarn bundle are generally distorted.} 
This distortion increases as the twist $\Omega$ of the bundle is increased and also grows with the radial coordinate of the filament from the center of the bundle, and can even result in concave and self-wrapping shapes~\cite{Greenvall2024}.
Since the distance between two filaments is determined by a straight line that must be orthogonal to the local tangent vectors of both filaments, this line cannot lie in the perpendicular cross-section of the twisted yarn, unless the yarn is not twisted ($\Omega = 0$) or both of the filaments simultaneously occupy the center of the yarn's core.
Therefore, the line of closest distance, when projected onto the yarn's perpendicular cross-section, has a reduced length that depends on the radial coordinates of the two filaments in question.
This projection gives rise to a radially-dependent metric tensor within the cross-sectional slice of the yarn, rationalizing the distortions of the fiber cross-sections.
It can be shown~\cite{Grason2015} that the induced metric is consistent with a space of positive Gaussian curvature $K_G$ that scales as $K_G \sim \Omega^2$.
This same Gaussian curvature frustrates hexagonal packing in two dimensions, resulting in a stress distribution within the yarn that is partially relieved through the introduction of 5-fold disclination defects~\cite{Bruss2012}.

\msd{So far, we have imagined that the fiber bundle has a continuous screw symmetry combining continuous translations and rotations along the fiber length.}
\msd{However, fibers towards the outer edge of the yarn must have larger arclengths than fibers in the core; the stretching energy of the filaments per yarn arclength then increases with radius.}
\msd{This causes filaments on the outside of the yarn to periodically \textit{migrate} towards the core}.
In doing so, they displace filaments within the yarn, which are then forced to migrate to the outer radii, as shown in \textbf{figure~\ref{fig:1}\textit{g}}.
Meanwhile, if the \msd{average linear} density of the yarn is expected to remain translationally-invariant, these migrations must be carefully coordinated~\cite{Morton1956}.
In order to maintain a translationally-invariant mass density of the yarn, the mass density cannot vary with the current radius of the migrating filament; within a small-twist approximation, this implies that the radial position of a migrating filament has the form $r(s) \approx R\sqrt{s/\lambda}$, where $s$ is the arclength coordinate of the yarn and $\lambda$ is the migration period~\cite{Hearle1965}.
While there are extensions to this idealized model~\cite{Treloar1965_1,Treloar1965_2}, much remains to be understood about the interactions between packing, migration, and contact mechanics.
Indeed, the simple assumptions of translational invariance and uniform density readily break down with the introduction of curvature~\cite{Atkinson2019}, with consequences for bundle mechanics that are still being explored through the language of nonlinear elasticity and gauge theory~\cite{Atkinson_2021}.

\subsection{Contact mechanics}

\msd{Pairs of yarn in contact locally impose forces and moments on each other, shaping their paths and dissipating energy through friction.}
Within the slender elastic curve model, these contacts can be interpreted as external local or distributed forces and moments \msd{along the yarn length}. 
The curve will then adapt its shape to minimize its elastic energy under these external constraints. 
The geometry of yarns in a textile is then principally determined by the arrangement of contacts imposed upon crafting.

\msd{Since pairs of yarn in contact cannot pass through each other they both experience a normal force that is simultaneously oriented transverse to the tangent vectors of both yarns at a shared contact point.}
\msd{Meanwhile, the existence of tangential forces implies the presence of frictional interactions.}
Friction dissipates energy when the contact is sliding, but can also statically trap tension within the yarn. 
For yarn mechanics, it is important to distinguish two types of frictional interactions: viscous friction and solid friction~\cite{Baumberger2006}. 
Viscous friction is typical of lubricated surfaces and is proportional to the relative speed between the contacting strands.
It then vanishes for a structure completely at rest. 
In contrast, dry friction holds tangential force until a certain threshold is reached and the contact starts sliding. 
Within the simplest Amontons-Coulomb formulation of dry friction, the threshold tangential force is proportional to the normal force and, once sliding, is independent of the sliding speed.
Which interactions are predominant may depend on the time scale, the type of yarn, or external conditions. 
Signatures of solid friction are found, for instance, in the hysteretic response upon cycling quasi-static tensile tests~\cite{Dusserre2015}, stick-slip instabilities emerging from sliding contacts~\cite{Poincloux2018PRL}, or the existence of multiple rest states~\cite{Crassous2024}. In contrast, viscous features are evidenced by slow creep-like responses, for instance, in the geometrical relaxation of the fabrics after fabrication~\cite{Munden1959}, or the stress relaxation of stretched fabrics~\cite{Matsuo2006}. 

In the case where the yarn is not made from one hard monofilament, the cross-section of the yarns will undergo significant deformation in the contact zones~\cite{durville2012contact, Grandgeorge2021}. 
Normal stresses in the contact zones will generally cause yarn to adopt an elliptical, or even ``stadium-shaped'' cross-section with flattened edges, as shown in \textbf{figure~\ref{fig:1}\textit{f}}.
These cross-section deformations may also be the source of dissipation, in particular if the yarn is made of multiple filaments in frictional contacts.
The interplay between filament packing and extreme cross-section deformations of spun yarn has been examined at some depth experimentally~\cite{hearle1969structural} but beyond recent investigations into twisted bundles described above in section~\ref{sec:fiber_packing}, little progress has been made in physical modeling.
Phenomenological yarn-yarn deformation mechanics models rely on power law scaling in the compressed volume of the fibers~\cite{kaldor8,singal2024programming}, motivated by van Wyk's studies on compressed wool~\cite{vanWyk1946}.

\section{TOPOLOGY AND SYMMETRY OF FABRICS}

\sonia{A fabric is a thin and flexible material made by intertwining yarns with respect to a particular manufacturing process.}
The 
\sonia{interlacing} process gives the fabric distinctive topological and symmetrical properties.
While a large variety of fabrics have been used throughout history, woven and knitted \sonia{structures exhibit periodicity along two transverse directions.}
This makes them analogous to two-dimensional crystals, which allows us to readily adopt the language and concepts of solid-state physics and crystallography.
We will focus on bulk descriptions of the material, leaving the interesting complexities of boundaries and interfaces, including \msd{selvedge in wovens and hemming in knits as well as the different structures resulting from the various cast-on and bind-off techniques,} for description elsewhere.
Instead, by focusing on the doubly-periodic bulk, we can simplify the structure of fabrics by focusing on a translational unit cell.

The unit cell representation also facilitates topological characterization of these fabrics.
\msd{Let $\mathcal{F}$ denote a topological fabric: a set of yarn paths embedded in the Euclidean thickened plane $\mathbb{R}^2 \times I \subset \mathbb{R}^3$, where $I$ is an interval representing the fabric's thickness (see the mathematical glossary in the margin for notation details)}. 
A primitive unit cell\footnote{The definition of a primitive or minimal unit cell may depend on the lattice being studied and which topological invariant is being considered.} $\tilde{\mathcal{F}}$ \msd{is a polyhedral region that} contains the minimal \sonia{number of yarn segments}
necessary to re-build the fabric \sonia{$\mathcal{F}$} through the repeated action of the \sonia{translation group,}
$G_{\rm tr}$, such that $\mathcal{F} = G_{\rm tr}\circ\tilde{\mathcal{F}}$.
\sonia{Then, by identifying the opposite faces of $\tilde{\mathcal{F}}$, we obtain a link in the thickened two-torus, $\mathbb{T}^2\times I$, defined as} the quotient space (or orbit space) $\mathcal{F}/G_{\rm tr}$ \sonia{of the fabric $\mathcal{F}$ under the maximal periodic lattice associated to $G_{\rm tr}$. 
This identification reduces the complexity associated with classifying the topology of an infinite collection of yarn and yarn crossings in $\mathbb{R}^3$ to the analysis of a few closed yarn paths, albeit in a space with a handle.
This makes the primitive unit cells particularly useful in the study of yarn linking topology since topological invariants can be constructed to classify fabrics~\cite{Grishanov2009I,Grishanov2009II, Fukuda2023, Diamantis2024}, up to doubly periodic equivalence~\cite{Diamantis2026}.}

\begin{marginnote}[0pt]
\entry{Group action $G\circ X$}{The set or space that results from the collection of images of $g$ applied to $X$, for all $g \in G$.}
\entry{Quotient space $X/G$}{A set of elements $e \in X/G$ 
\sonia{such that for each $e$ there exist no element $e' \in X/G$ equivalent to $e$ under the action of $G$.}}
\entry{Product space $X\times Y$}{Space whose elements are ordered pairs $(x,y)$ obtained from two different spaces $X$ and $Y$, with $x\in X$ and $y \in Y$.}
\end{marginnote}



Two-dimensional tilings of the plane can be completely classified using only 17 symmetry groups, the wallpaper groups~\cite{schwarzenberger:1974, schattschneider:1978, armstrong:1988}.
These groups are essential for understanding the structure and response of 2D crystals~\cite{Dresselhaus2025AnomalousTensorial2D} and, as their name implies, are found in works of art, as exemplified by the tilings of M.C.~Escher~\cite{schattschneider2010mathematical}.
However, fabrics can have symmetries through their thickness, requiring description using \emph{sub-periodic symmetry groups}, where the number of directions with translational periodicity is less than the embedding dimension.
Of these, fabrics are described by the 80 layer groups (2D periodicity in 3D)~\cite{Liu2018,mahmoudi2025}.


\subsection{Wovens}
Most woven fabrics are \textit{biaxial}, with yarn running in perpendicular directions known as warp and weft.
When weaving on a loom, many parallel ``warp'' yarns are typically oriented to point towards the weaver who uses a shuttle to run a working ``weft'' yarn back and forth across the warp yarns. 
In modern industrial looms, instead of shuttles, many other weft insertion processes are now used, including rapier, air-jet, and water-jet processes for production efficiency, although the process remains largely the same. 
In these looms the warp yarns are lifted up or lowered in a patterned sequence through the use of heddles, threads or cords containing eyes through which the warp yarns are threaded, forming an opening called the ``shed" in which the weft is inserted, resulting in an arrangement of over and under-crossings that can vary from row to row. 
The Jacquard loom, due to its ability \sonia{to control each heddle independently and} to re-configure the heddle lifting sequence using punch cards, is often considered the first computer \cite{postrel2020fabric}. 

\begin{figure}
\includegraphics[width=\textwidth]{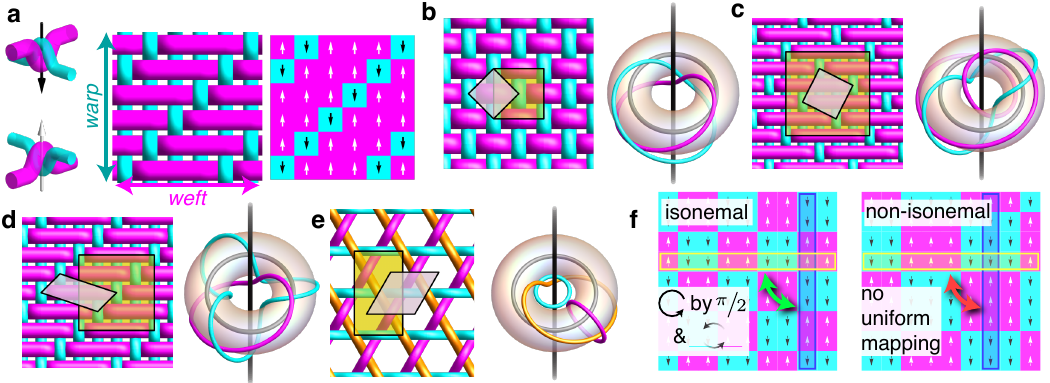}
\caption{Symmetries and topology of woven fabrics. (a) Mapping from order of yarn crossings to a two-state spin lattice, illustrated by a (3,1)-twill pattern. Typical woven patterns shown include plain weave (b), \msd{(5,3)}-satin (c), (3,1)-twill (d), and the Kagome weave (e). Each of these woven patterns has a weft-oriented translational unit cell shown in yellow and a primitive unit cell shown in white. The quotient space of the primitive unit cell, $\mathbbm{T}^2\times I$, is shown with closed curves for each yarn\msd{, represented as links in $S^3$}. (f) Examples of an isonemal $2\times 2$ basket weave, where a uniform transformation can be used to exchange any two yarns without changing the fabric pattern, and the non-isonemal $3\times 3$ basket weave.}
\label{fig:2}
\end{figure}

We will refer to an ``over-crossing'' (resp., ``under-crossing'') as a region of the fabric where \emph{the weft yarn passes over \sonia{(resp., under)} a warp yarn}. 
Due to the apparent grid-like nature of the biaxial weaves, \sonia{the over/under-crossing arrangement has long been encoded on rectangular grids, such as that shown in \textbf{figure~\ref{fig:2}\textit{a}}, which can also be viewed as binary matrices} \msd{or $\mathbb{Z}_2$ spin lattices}.
\msd{The net spin of a $(p,q)$-twill (see margin note) is determined by the ratio of the difference in up and down-crossings $p-q$ to the total number of crossings $p+q$; the (3,1)-twill pattern shown above has a net spin of $+1/2$.} 
By contrast, the plain weave shown in \textbf{figure~\ref{fig:2}\textit{b}} consists of an equal arrangement of over and under-crossings and 
\sonia{can be seen as} a $(1,1)$-twill; this structure has a distinct antiferromagnetic arrangement of crossings.
Note that of the patterns shown in \textbf{figure~\ref{fig:2}\textit{b-e}}, only the plain weave and its generalizations in \textbf{figure~\ref{fig:2}\textit{f}}\sonia{, called basket weaves,} have mirror symmetries.



Given the orthogonal arrangement of yarns in biaxial weaves, it is natural to define the base repeat pattern using a translational unit cell whose edges align with the warp and weft directions; these are shown in \textbf{figure~\ref{fig:2}\textit{b-e}} in yellow.
If one considers the primitive unit cell (shown in white) of these specific structures, the resulting lattice directions generally depart from the warp and weft directions.
As shown in the right panels of \textbf{figure~\ref{fig:2}\textit{b-e}}, the use of a primitive unit cell can even reduce the number of independent yarns to two for the biaxial examples and three for the Kagome triaxial weave.

The topology of each of these patterns is invariant under smooth deformations of the yarn paths\sonia{, namely under ambient isotopy in the thickened torus.}
\sonia{Unlike classical links in the three-sphere $S^3 =\mathbb{R}^3 \cup \{\infty\}$, links in $\mathbbm{T}^2\times I$ contains curve components that do not shrink to points, since the torus} 
has two directions or ``generators'' around which a 
\sonia{component} can wrap.
In the case of the Kagome weave (also called the ``mad weave'') in \textbf{figure~\ref{fig:2}\textit{e}}, the magenta and cyan 
\sonia{curves} each wrap different generators of the torus once and therefore cannot be shrunk to a point, even if the yellow curve was removed.
\sonia{By placing $\mathbbm{T}^2\times I$ in $S^3$,} the generators of the torus may be regarded as providing two additional curves \sonia{to represent the link in $S^3$}, one that is in the form of a closed \sonia{curve ``parallel'' to a  longitude}
and the other that extends through the axis of revolution of the torus but has its antipodal ends at $\pm \infty$ identified to form a second closed 
\sonia{curve ``parallel'' to a meridian} \cite{Morton2009}.
Therefore, the topology of the plain weave, satin, and twill patterns shown in \textbf{figure~\ref{fig:2}} can each be described through the linking of four curves \sonia{in $S^3$} (two yarn, two torus generators), whereas the Kagome weave links five curves \sonia{in $S^3$} (three yarn, two torus generators).
It should be noted, however, that there is some ambiguity through the choice of unit cell. \sonia{More precisely,} a different choice of 
\sonia{basis of the plane corresponds to a finite sequence of full twists (Dehn twists) of the torus quotient and} can change how \sonia{the same} yarn 
\sonia{components} wrap the generators 
without changing the overall topology of the resulting fabric~\cite{Grishanov2007, Fukuda2023,Diamantis2026}.
The choice of unit cell for the biaxial structures considered here requires that each closed curve traverses both generators at least once, and the number of wraps of the warp yarn around the minor axis of the torus maps to the number of warp yarns between two under-crossings in neighboring rows: one for the plain weave (i.e., (1,1)-twill), two for satin, and three for the (3,1)-twill.

\begin{marginnote}[0pt]
\entry{Warp}{In the context of weaving, the lengthwise collection of yarns in the fabric as woven.  The term is also used to denote the lengthwise direction of the fabric.}
\entry{Weft}{In the context of weaving, the collection of yarns inserted across the fabric width (perpendicular to the warp); also refers to the widthwise direction.}
\entry{Crimp} {Waviness or undulation of a yarn in a woven fabric.}
\entry{$(p,q)$-twill}{A woven pattern having $p$ over-crossings, followed by $q$ under-crossings with subsequent rows shifting the pattern by one crossing~\cite{Liu2018}.}
\entry{$(n,s)$-satin}{An $n\times n$ woven pattern with one under-crossing in each row and $s$ over-crossings between the under-crossings in neighboring rows.}
\entry{Isonemal}{Property by which any two yarns can be exchanged by an operation that maintains the symmetry of the fabric's pattern.}
\end{marginnote}

Beyond the usual symmetry groups that characterize 2D materials, including woven fabrics~\cite{Liu2018}, there exists an additional ``isonemal'' (\textit{iso-} meaning ``same,'' and \textit{nemal} meaning ``thread'') symmetry that describes the ability for two yarns to be exchanged without disruptions to the pattern.
In the most general setting, this symmetry requires that warp and weft yarns must be able to be exchanged.
\sonia{Note however that since the warp yarns are under different mechanical loading conditions than the weft yarns and are often made from different materials, such idealized symmetries may not fully capture realistic weaves.}
The yarn exchange can involve any uniform (homogeneous) operation consisting of isometries of the fabric~\cite{Grunbaum1980,Zelinka1984}; this means that flipping all over-crossings to under-crossings is allowed, albeit limited to the case where there are the same number of over and under-crossings (e.g., a $(p,q)$-twill with $p=q$).
Importantly, all regular $(p,q)$-twills are isonemal~\cite{Grunbaum1980}.
Satin structures \msd{(see glossary definition)}, on the other hand, are not all isonemal.
The \msd{(5,3)}-satin\footnote{\msd{Note that the (5,3)-satin and the (5,2)-satin are chiral opposites.}} in \textbf{figure~\ref{fig:2}\textit{c}} is isonemal, whereas a (7,2)-satin is not~\cite{Grunbaum1980}.
For triaxial weaves, the Kagome in \textbf{figure~\ref{fig:2}\textit{e}} is the simplest isonemal pattern~\cite{Liu2018}.
\textbf{Figure~\ref{fig:2}\textit{f}} shows a comparative example of two basket weaves.
While the $2\times 2$ basket weave is isonemal, the $3\times 3$ basket weave (as well as $n\times n$ for $n \geq 3$) is not.
Here, the difference is clear: each crossing in the $2\times 2$ basket weave neighbors two over-crossings and two under-crossings, whereas the $3\times 3$ basket weave also has crossings that have all under-crossing or all over-crossing neighbors, resulting in families of yarns that cannot be exchanged by an isometry.
This hints at a deep connection between the isonemal property and layer group symmetries~\cite{Roth1993}.


\subsection{Knits}

\begin{figure}
\includegraphics[width=\textwidth]{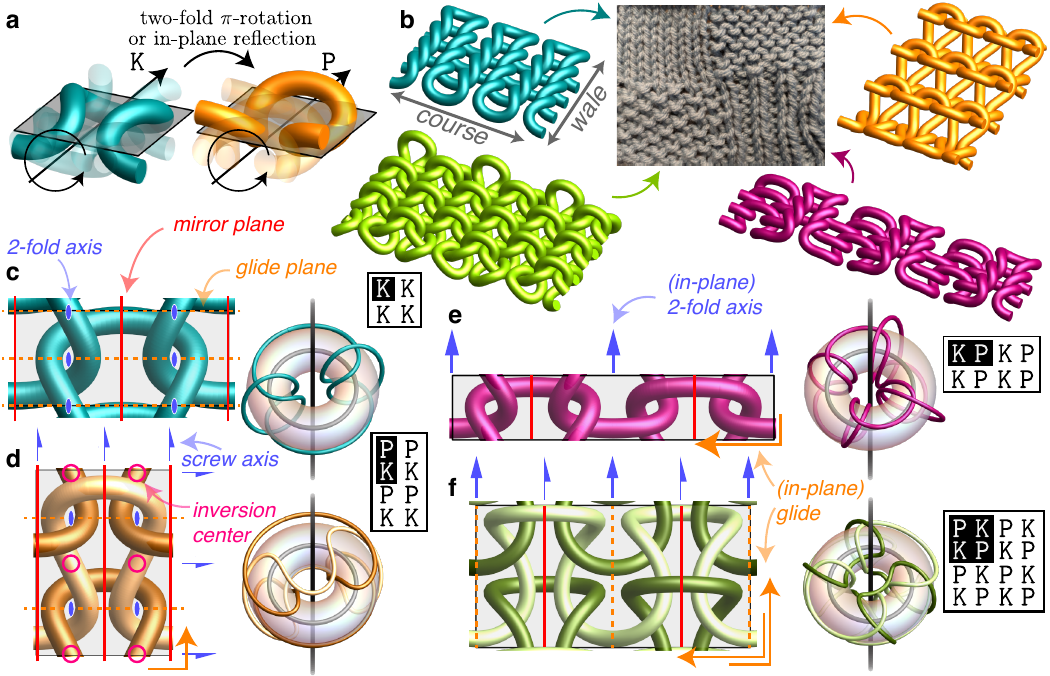}
\caption{Structure of knits. (a) Knit (\texttt{K}) and purl (\texttt{P}) stitches are related by either a $\pi$-rotation about an axis in the plane of the fabric or by a mirror operation through the plane of the fabric. (b) Typical patterns of doubly-periodic knitted fabric, including (clockwise from the upper left): plain knit, garter, rib, and seed. The primitive translational unit cells are shown with crystallographic symbols representing elements of the layer symmetry groups, along with the link structure \msd{in $S^3$}, for plain knit (c), garter (d), rib (e), and seed (f).}
\label{fig:3}
\end{figure}


Knitting is fundamentally different from weaving, both on a structural level and in its creation: knitting machines use hooks and hand-knitters just require a pair of \msd{long} needles.
While the unit cells of weaves are \msd{composed of} crossing strands, the unit cells of knitted fabrics are linked loops. Here and throughout this review, we focus on \emph{weft} knitting \msd{(see margin note)}. 
The linking process for weft knitting occurs in three stages:
\begin{enumerate}
    \item A row of connected loops is made by tying a series of slip knots (casting on).
    \item For each of the loops in the first row, a new loop is created and linked to the existing loop. After completion of this process for every loop in the first row, the process is repeated until the fabric achieves the desired length (knitting).
    \item The loops of the final row are combined into a knot that ensures the fabric does not unravel (casting off).
\end{enumerate}
In weft knitting, there is no need for warp yarns: in its most basic form, only one yarn is used at a time. 
Thus, instead of distinguishing fabric directions by warp and weft, the term \textit{course} is used to describe the row-wise direction along which arrays of loops are introduced into the working yarn, and the term \textit{wale} is used to describe the column-wise direction where loops are brought through each other.
To join two loops, either yarn may be pulled from the back of the fabric, resulting in a \textit{knit stitch}~\texttt{K}, or from the front, resulting in a \emph{purl stitch}~\texttt{P}.
The underlying structure of these stitches is identical, depending only on the orientation of the fabric; a \texttt{K}-stitch may be reflected or transformed by a $\pi$-rotation of the fabric about the wale axis to create a \texttt{P}-stitch, as shown in \textbf{figure~\ref{fig:3}\textit{a}}.
While there are other types of stitches, such as twisted versions of the \texttt{K} and \texttt{P} stitches, most of the common knitted fabrics are made from these two. 
As a result, much like the two states of crossings in woven fabrics, it is possible to enumerate the \texttt{K}/\texttt{P}-based two-periodic knitting patterns using a 
\sonia{rectangular} grid.
The canonical knitting patterns are shown in \textbf{figure~\ref{fig:3}\textit{b}}: \emph{plain knit} fabric (or ``stockinette'' for hand-knitters, ``jersey'' for machine-knitters) consists of a lattice of all \texttt{K}-stitches (or all \texttt{P}-stitches when viewed from the reverse side); \emph{garter} fabric alternates between courses of \texttt{K}-stitch and courses of \texttt{P}-stitch; \emph{rib} fabric alternates between wales of \texttt{K}-stitch and wales of \texttt{P}-stitch; \emph{seed} fabric has a checkerboard arrangement of \texttt{K}s and \texttt{P}s.
Notably, not only do these fabrics have distinct textures, but they also have remarkably distinct mechanical properties~\cite{singal2024programming,Ding2024}.

\sonia{Similar to woven fabrics, one can study the symmetries of knitted patterns through their unit cells.}
The unit cell for plain knit fabric, shown in \textbf{figure~\ref{fig:3}\textit{c}}, consists of a single yarn segment.
Each of the course-wise connecting arcs bridge a mirror plane, represented by a red line.
There are also sets of two-fold rotation axes that are oriented perpendicular to the plane of the fabric; when combined with the mirror planes, these rotation axes give rise to glide planes that cut through the fabric and are oriented along the course direction.
All of these are consistent with the symmetry group pma2, which is also one of the 17 wallpaper groups\sonia{, up to thickness }.
By contrast, garter stitch has a unit cell containing a pair of \texttt{K} and \texttt{P} stitches.
Since \texttt{K} and \texttt{P} stitches are related by reversing the fabric, this gives rise to a glide symmetry through the plane of the fabric, as shown in \textbf{figure~\ref{fig:3}\textit{d}}.
When combined with the mirror planes, the result is a set of in-plane screw axes.
Finally, one set of the two-fold rotation axes is replaced with a set of inversion centers.
Unlike plain knit fabric, garter fabric contains symmetries through the thickness of the fabric and is therefore described by a layer group, pbma, which is not \sonia{a direct extension of} one of the wallpaper groups~\cite{OKeeffeTreacy2022Isogonal,mahmoudi2025}.

Rib and seed fabrics, shown in \textbf{figure~\ref{fig:3}\textit{e-f}} have \texttt{K} and \texttt{P} stitches that neighbor each other in the course direction.
\ejd{Thus, compared to plain knit and garter fabrics, one mirror plane is lost and replaced by in-plane two-fold axes that serve as mechanisms to convert between \texttt{K} and \texttt{P} stitches}
However, there is an additional in-plane glide symmetry along the course direction, with the seed fabric possessing a second glide symmetry along the wale direction.
The seed fabric has an additional set of in-plane screw axes that are aligned with the remaining mirror planes.
The rib fabric is described by layer group pm2a, and the seed fabric by layer group cm2e.
Note that neither of these fabrics has a two-fold axis out of the plane, unlike the plain knit and garter, or any of the woven fabrics that we have considered in this review; \msd{rib and seed fabrics are therefore polar and may be of interest in the context of polar mechanical metamaterials~\cite{Pishvar2020}}.

\begin{marginnote}[0pt]
\entry{Weft knitting} {Knitting fabrication process in which loops are formed successively along the course direction.}
\entry{Warp knitting} {Knitting fabrication process in which loops are formed in parallel along the wale direction, linking neighboring wales. 
\entry{Course}{In the context of knitting, a row of loops in the fabric, oriented parallel to the fabric's width.}
\entry{Wale}{In the context of knitting, a column of loops in the fabric, oriented parallel to the fabric's length.} 
\entry{Stitch} {Individual or pattern of loops in a knitted fabric.}
}
\end{marginnote}

The topology of knitted fabrics is also distinct from that of woven fabrics. \sonia{One of the main structural difference comes from the fact that all yarn paths run through parallel directions in knitting, while at least two transverse directions are needed for weaving. Moreover, a single yarn can self-cross in knitting, while no yarn crosses itself in weaving}
\sonia{For the plain knit and rib stitch, a primitive unit cell can be formed by a single knot, while those of garter and seed fabrics needs an additional knot component}. 
The structure of knits has been described by topological frameworks in~\cite{Grishanov2007} and~\cite{Markande2020}; the latter work finds a topological classification for all structures that can possibly be knit. 


\section{FABRIC MECHANICS AND APPLICATIONS}

Although yarns and fabrics have been used for millennia, quantitative descriptions of their mechanics still face enormous difficulties: the emergent behavior of fabrics is generally nonlinear and anisotropic, with effective bulk properties that depend on geometric and topological features appearing at the same length scale as the fabric thickness. 
The manufacturing process of ordered textiles fixes their topology, creating yarn crossing constraints that are not localized in the yarn path; this contributes greatly to computational cost in yarn-level descriptions~\cite{hu2024knitted}. 
These topological constraints are difficult to elide because they determine the structural integrity of the fabric; in knits, force is largely transmitted through yarn contact points between locally non-contiguous segments of yarn. 
Additionally, the length scale of individual stitches or weaving motifs is typically comparable to the length scale of the yarn thickness.

\begin{figure}
\includegraphics[width=\textwidth]{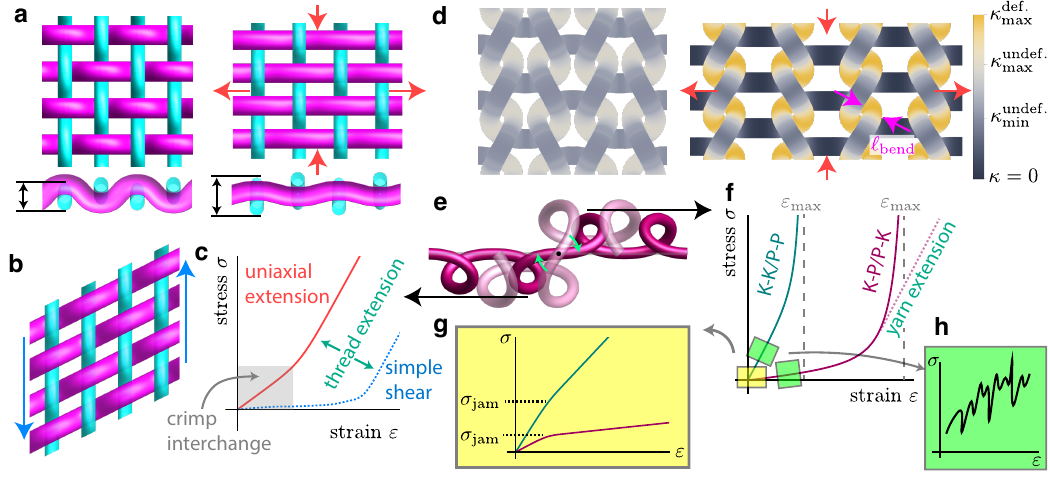}
\caption{Mechanics of woven and knitted textiles. (a) A plain weave is shown in its undeformed reference configuration, with side profile showing the crimped structure of both warp and weft yarns. A weft-wise uniaxial extension flattens the weft yarns, increasing the crimp of the warp yarns, resulting in a positive Poisson effect. (b) A simple shear deformation without significant deformation of the yarns. (c) Stress-strain plot showing the crimp interchange under uniaxial deformation and the soft shear mode, both of which strain-stiffen upon extension of the yarns. (d) Plain knit fabric under a similar uniaxial extension along the course direction, with curvature distribution $\kappa$ shown as a color gradient map; the bending lengthscale $\ell_{\rm bend}$ is increasingly localized to yarn clasp regions. (e) The \texttt{K}-\texttt{P}/\texttt{P}-\texttt{K} structure exemplified by the rib stitch has a soft deformation mode associated with rotation of the \texttt{K}/\texttt{P} boundary. (f) Typical stress-strain curves for fabrics whose deformations involve stretching \texttt{K}-\texttt{K}/\texttt{P}-\texttt{P} boundaries (teal) and \texttt{K}-\texttt{P}/\texttt{P}-\texttt{K} boundaries (purple). These curves stiffen and asymptote at a maximum strain $\varepsilon_{\rm max}$ for inextensible yarn but yield to a finite slope if the yarn is extensible (dashed line). (g) Inset shows an initially stiff, ``jammed'' response for low stress ($\sigma < \sigma_{\rm jam}$) for tightly-knitted fabric. (h) Crackling behavior resulting from stick-slip events can be seen as the fabric is stretched.}
\label{fig:4}
\end{figure}

Nevertheless, much work over the past century has empirically characterized fabric mechanics across various industrial manufacturing methodologies and yarn materials. 
Theoretical work in this area typically falls into complementary categories: finite-element volumetric yarn modeling, and approximate physical simulation including both approximate yarn-level computations of textile mechanics~\cite{popper1966theoretical, kaldor8, durville2010simulation, cirio_yarn-level_2017, breen1994predicting, eberhardt1996fast, kyosevFEM2012} as well as fabric-level approximations that treat swatches of textiles as discretized two-dimensional surfaces modeled as thin elastic shells~\cite{weil1986synthesis, feynman1986modeling, shanahan1978characterizing, narain2012adaptive, tamstorf2015smoothed}. 
With substantial increases in computational power, recent work has also focused on database-driven approaches, where different physical models or textures are used in different regimes of deformation, and hybridized computations of yarn- and fabric-level models~\cite{yuksel2012stitch, casafranca2020mixing, Sperl2020, sperl_estimation_2022}.

\subsection{Wovens}



Biaxial woven structures \msd{have} extremely anisotropic in-plane mechanics~\cite{cao2008characterization}. 
If one tries to pull in the warp or weft direction, parallel to the yarns, \msd{the yarns are put under primarily tensile or compressive loads~\cite{grosberg1966_I,kawabata1973_I}, resulting in great in-plane stiffness.}
The \emph{deformation} response is determined by the geometry of the serpentine path that perpendicular yarn segments are forced to adopt when they are woven under tension \msd{(known as \textit{crimp})}.
Under conditions of mechanical equilibrium without external stress, the crimp adopts a characteristic spacing and curvature for both warp and weft yarns.
Under uniaxial extension, for example along the weft direction as shown in \textbf{figure~\ref{fig:4}\textit{a}}, the yarns aligned in the extended direction straighten, which in turn forces the perpendicular fibers to increase the amplitude of their out-of-plane undulations, resulting in a positive 2D Poisson effect.

If one pulls in a direction not aligned with the yarns, the yarns can rotate around their point of contact with only minute deformations, resulting in soft shear modes~\cite{lo2002shear}, illustrated in \textbf{figure~\ref{fig:4}\textit{b}}. 
As shown in \textbf{figure~\ref{fig:4}\textit{c}}, the elastic stiffness is dominated by the uniaxial response, which is determined by crimp interchange for low strains, $\epsilon \lesssim \mathcal{O}(10^{-1})$.
For higher strains, the fabric rapidly strain-stiffens, with the modulus determined by the extensional stiffness of individual yarns.
Despite the near-negligible cost of uniform shear deformations, frictional contacts play a significant role, leading to highly hysteretic responses~\cite{grosberg1966_V,kawabata1973_III}. 

Weaves are very slender sheets so their out-of-plane deformations much softer than their in-plane deformations~\cite{grosberg1966_IV}. 
\msd{If the crossed yarns of a biaxial weave have an approximately rectangular structure, then the curvatures that are compatible with the in-plane mechanics of the weave must keep the length of the rectangles’ sides fixed (inextensibility of the yarns) but can allow the angles to freely vary (shear soft mode).}
This leads to a formal relation between the low-energy deformed shapes of woven fabrics and the mathematics of Chebyshev nets or gridshells~\cite{tchebychev1878coupe,ghys2011coupe}.
In-plane patterning can be designed such that the fabric will wrap complex shapes with single or multiple layers of woven fabrics, which is of utmost importance for industrial applications of textiles-reinforced composites \cite{boisse2018bending,chen2011overview}.

The mechanical response of biaxial weaves is thus primarily determined by the inextensibility of the yarns and possible rotation at the \msd{yarn-yarn} contacts. 
Consequently, biaxial weave mechanics is only marginally sensitive to the type of weave (plain, twill, etc.).
The type of weave, however, is the primary factor influencing the touch feeling of fabrics, a crucial property for any garment application~\cite{bertaux2007relationship}.
However, if one relaxes the strictly orthogonal warp/weft structure of biaxial weaves and allows for the introduction of defects, then the 3D structure and mechanics can be greatly altered.
This is illustrated by the case of woven baskets, where angle deficits enable the creation of positive Gaussian curvature defects---namely, corners---which are significantly stiffer than the surrounding biaxially-woven structure~\cite{Tu2025}.
Woven textiles are an ideal playground for designing 3D structures through the use of disclination defects~\cite{Seung1988,Callens2018}.

Triaxial textile weaves have a long history as a traditional crafting method, notably in basket weaving. 
Adding a series of yarns in a third direction considerably stiffens the structure, as the shear mode now also involves yarn stretching. 
Triaxial weaves are usually crafted from ribbons instead of threads with circular cross-sections to allow the structure to help maintain a high cover factor with minimal out-of-plane bending of the yarns at the crossings. 
Ribbons can only bend out of their plane, imposing strict in-plane geometrical constraints: ribbons must follow the geodesics of the surfaces they embed. 
\msd{Consequently}, triaxial weave \msd{patterns determine the apparent Gaussian curvature of the fabrics and thus the 3D shape~\cite{baek2021smooth}}. 
Weaving the ribbons into hexagonal patterns generates flat surfaces, but introducing topological defects like pentagons or heptagons leads to dome- and saddle-shaped surfaces, respectively~\cite{ayres2018beyond}. 
\msd{Additionally, the prescription of Gaussian curvature in turn dramatically affects the mechanical response of the woven surface~\cite{poincloux2023indentation}}.

\subsection{Knits}

Unlike woven fabrics, knits tend to have soft in-plane and out-of-plane mechanical responses due to highly curved yarn paths, which allow them to significantly deform without substantial extension of the yarn along its length.
Large yarn curvatures (typically at a similar length scale to their thickness) \msd{lead to emergent nonlinear behavior in stress-strain constitutive relations}~\cite{Audoly2010}, and knits display an abundance of yarn contacts at nontrivial angles, requiring friction forces to be modeled for quantitative accuracy even at low tensions~\cite{miguel2013modeling}.
In addition to generalized continuum approaches also used for woven fabrics \cite{yeoman2010constitutive}, simulation methods for knits include multiscale constitutive modeling relating the complex three-dimensional structure of knitted loops with fabric-scale deformations~\cite{Liu2017, dinh2018prediction, tekerek_experimental_2020,narain2012adaptive,tamstorf2015smoothed} as well as yarn-level simulations~\cite{kaldor8,kaldor_efficient_2010,cirio_yarn-level_2017,singal2024programming,Ding2024}.
Despite the complexity involved in computational models, there is a long history of using geometric models to establish basic results governing stress-strain relationships and fabric stiffness~\cite{postle_24analysis_1967,Postle2002,RAMGULAM201148,semnani_new_2003,htoo_3-dimension_2017,abghary1026,ru_modeling_2023,Poincloux2018PRX,singal2024programming}.

To outline a basic formulation of this problem, consider the relaxed configuration of an individual stitch, which is controlled by the tension under which the yarn was knit (or the length of yarn per stitch) as well as boundary \msd{conditions arising from boundary stitch structure and mechanical constraints}.
Using the yarn's bending modulus $B$ and the tension $T$, one can define a \textit{bending length} $\ell_{\rm bend} \equiv \sqrt{B/T}$, which characterizes the \msd{size of the region over which curvature varies.} 
If the fabric is knit under low tension, the stitch curvature \msd{$\ell_{\rm bend} \sim L$, where $L$ is the length of yarn per stitch, representing an upper characteristic lengthscale.} 
\msd{If the repeat unit of a pattern involves multiple stitches, then $L$ can be the length of yarn within a unit cell.}
If the fabric is knit under high tension or an external stress is applied, the bending length decreases, approaching a \msd{lower} characteristic lengthscale of the stitch, namely the yarn radius $r$.
As shown in \textbf{figure~\ref{fig:4}\textit{d}}, the curvature transitions from being delocalized throughout the stitch to being localized to the cross-over \msd{(or ``clasp'')} regions.
Moreover, the length $\ell_{\rm bend}$ can serve as a type of interaction lengthscale between stitches: for large $\ell_{\rm bend}$, the pattern of stitches determines the mechanical response of the knit, whereas for small $\ell_{\rm bend}$, the mechanical response is determined by the stiffness of the yarn and the contact mechanics at the clasp regions between pairs of yarn segments~\cite{popper1966theoretical}.
The pattern dependence has a rather simple rule: if two neighboring stitches are different (\texttt{K}-\texttt{P} neighbors) then there is a soft deformation mode that exists due to a near-rigid rotation of the connecting segment, as shown in \textbf{figure~\ref{fig:4}\textit{e}}; if the neighboring stitches are the same (\texttt{K}-\texttt{K} or \texttt{P}-\texttt{P} neighbors) then the stiffness is roughly an order of magnitude larger~\cite{singal2024programming}, as shown schematically in the stress-strain plot in \textbf{figure~\ref{fig:4}\textit{f}}.
This model confirms that the basic trends observed for knitted fabrics come from stitch geometry: plain knit is uniformly the stiffest and while rib has an extreme soft deformation when stretched in the course direction, garter and seed are the softest when stretched in the wale direction~\cite{singal2024programming, Ding2024}.

Beyond the initial linear regime, knitted fabrics have an extended nonlinear elastic response where they smoothly strain-stiffen, exhibiting the usual J-shaped curve typical of amorphous fiber networks found in biological tissue~\cite{Storm2005}.
For inextensible yarn, this strain-stiffening eventually asymptotes at a maximum strain cutoff $\varepsilon_{\rm max}$, as shown in \textbf{figure~\ref{fig:4}\textit{f}}, beyond which the yarn fails; if the yarn is extensible, then this asymptote is not reached and a second, high-stress linear regime is reached, where the stiffness is governed by the extensional modulus of the yarn~\cite{Ding2024}.
The divergence of the stress as $\varepsilon \to \varepsilon_{\rm max}$ in fabrics made from inextensible yarn follows the form $\sigma \sim (\varepsilon_{\rm max} - \varepsilon)^{-2}$, which can be derived as an asymptotic limit of an inextensible elastica model of yarn, where curvature is increasingly localized to infinitesimal regions under increasing tension~\cite{singal2024programming}.
Notably, this same asymptotic force law is obeyed by DNA and other wormlike polymer chains as they are stretched to their maximum length, \msd{although in the polymer case, this is due to} the reduction in conformational entropy under high tension rather than curvature localization~\cite{Marko1995}.

\subsubsection{Analogies with granular mechanics}

Besides sharing common features with fiber networks, \msd{variability in yarn-yarn contact regions can result in some surprising similarities with particulate granular systems.} 
An additional stiff regime can be observed at low stresses when the knitted fabric is knit in a tight or ``jammed'' configuration.
This jammed regime is dominated by a combination of contact mechanics and large bending stresses, but softens to the usual pattern-dependent regime after a sufficiently large applied stress $\sigma \gtrsim \sigma_{\rm jam}$ as shown in \textbf{figure~\ref{fig:4}\textit{g}}.
\msd{The threshold stress for the jammed regime} has been shown to smoothly vanish ($\sigma_{\rm jam}\to 0$) as the length of yarn per stitch is increased, indicating that it is a density or volume-fraction controlled transition, similar to granular jamming~\cite{Gonzalez2025}.

Interestingly, the analogy with granular materials also extends to the observation of \textit{crackling noise}, which can be found in the force fluctuations measured in earthquakes and amorphous solids.
Such materials respond to external loads by local yielding events, which redistribute accumulated stresses to neighboring regions, increasing the probability that these regions also yield. 
Deformations then proceed via avalanching yield events with typical power-law statistics \cite{nicolas2018deformation}.
As a loose-knitted fabric is slowly stretched, the stress-strain curve starts to exhibit small-scale stress drops of various sizes, as illustrated in \textbf{figure~\ref{fig:4}\textit{h}}.
These stress drops are caused by local stick-slip instabilities at the yarn contacts, which redistribute stress anisotropically, creating extended slip lines. Even though the knit structure is regular, frictional instabilities create analogs of local yielding events in amorphous solids that avalanche into large-scale events with power-law statistics~\cite{Poincloux2018PRL}.

This combination of fiber and granular mechanics places knits and wovens as examples within the emerging field of ``elastogranular materials,'' which involves the overlap of slender body mechanics, granular packing, and tensegrity~\cite{Guerra2022,Guerra2023,Dreier2025}.

 \subsubsection{Hysteresis and memory of mechanical response}
The tensile mechanical response of knitted fabric shows significant hysteresis, \msd{resulting from} the dissipative nature of the deformation process. 
Attempts to model and predict this hysteretic behavior \msd{invoke} either viscoelasticity~\cite{Matsuo2009HysteresisKnitted} or dry friction~\cite{Dusserre2015}. 
\msd{More recent} work hypothesized that this effect comes mostly from plastic deformation with a small viscoelastic effect, \msd{rationalized by an observed independence of hysteresis on applied strain rate}~\cite{dresselhaus2025knittingmemory}.
\msd{This latter study} also observed and modeled return point memory in knitted fabrics, \msd{drawing on an analogy with history effects in ferromagnets}~\cite{sethna1993}. 
In this type of memory, the fabric ``remembers" the load at the maximum strain reached during the tensile test (the return point). 
The fabric ``forgets" the return point when that maximum strain is exceeded and instead remembers the new return point. 

\msd{A similar memory effect results in the relaxed (i.e.~unloaded) shape of knitted fabrics adopting different aspect ratios.}
The multiplicity of rest shapes poses important challenges for industry in predicting the surface of fabric from a given length of yarn, fabrication parameters, and post-processing~\cite{Munden1959}. 
\msd{This is caused, in part, by} friction at the contacts, which traps internal stresses and prevents the fabric from relaxing to its minimum energy state~\cite{Crassous2024}. 
Depending on the deformation history, the \msd{fabric} can get stuck at various aspect ratios, thereby encoding memory in the knit's rest shape.

\subsubsection{Anisotropic fracture in knits}

The local structure of yarn paths within a knit fabric also leads to highly anisotropic and heterogeneous patterns under mechanical failure~\cite{faulconnier2026laddering}. When a yarn tears locally in a knit fabric, biaxial \msd{stress in} the fabric may propagate \textit{runs} or \textit{ladders} from the initial breakage site, creating defects in the loop structure. Under biaxial \msd{stress}, yarn breakage allows for two primary modes of defect propagation: (1) \textit{reptation} of the newly created free ends, reducing the total number of loops in a row and thereby releasing tension in the $y$ direction; and (2) loops that are no longer topologically constrained by a vertically neighboring loop may pull taut horizontally, releasing tension in the $x$ direction. Due to the relative amount of friction required to be overcome in each mode of propagation, loop defects typically travel much farther in the wale ($y$) direction than the course ($x$) direction, leaving long ladder-like scars in the fabric. 
\ejd{The topology of the knit pattern restricts how a fracture propagates. Initial damage, typically a yarn breaking missed loop, modifies the topology of the knit, creating an effective \textit{topological defect}. Damage propagation, however, conserves topology, limited to Reidemeister moves of the yarn~\cite{shimamoto2025topological}. }
Recent work has begun to quantify fracture propagation in weft knit composites~\cite{liu2024composite}, although this phenomenon has been seldom studied outside of sartorial or forensic contexts~\cite{toyoma1980some, dann2012tearing}. Warp knits, due to their yarns' vertical orientation, resist laddering and are often used in applications requiring wear resistance.

\begin{marginnote}[0pt]
\entry{Reptation}{Polymer physics term describing ``snake-like'' motion or displacements of a curve (e.g.~polymer or yarn) along its tangent.}
\end{marginnote}


\begin{figure}
    \includegraphics[width=\textwidth]{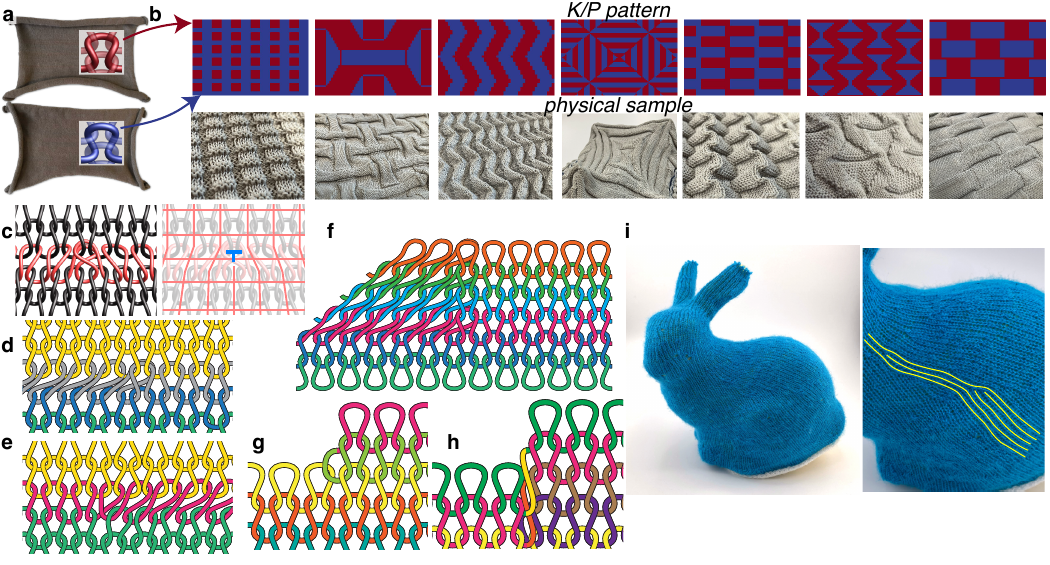}
    \caption{(a) Plain knit with \texttt{K}-side shown has the first and last rows/courses curling up (top); the \texttt{P}-side has the first and last columns/wales curling up (bottom). (b) Larger-scale patterns (top) created by combinations of \texttt{K} (red) and \texttt{P} (blue) stitches with physical samples (bottom) showing the relaxed 3D structures (figure reproduced from~\cite{niu2025geometric}). (c) A row decrease created with a knit-two-together (\texttt{k2tog}) operation results in an edge dislocation. A row decrease (d) or increase (e) near the end of the fabric has an asymmetric effect. (f) A line of row decreases can create a tilted domain. ``Short rows'' can be used to shape fabric boundaries (g) or create disruptions in lattice structure by advancing part of it in the wale direction. (i) A 3D-knit pattern of the Stanford bunny shows how short rows, decreases, and increases can be used to create curvature (photos and model courtesy of James McCann, Carnegie-Mellon University); a close-up view with example edge dislocations highlighted is shown on the right.}
    \label{fig:5}
\end{figure}

\subsubsection{Curling of knit fabric}

Without external forces or postprocessing, weft knitted fabrics also have a bulk tendency to curl dramatically at the fabric scale \cite{knittel2015self, pavko2017multifunctional, tajiri2025curling, niu2025geometric}. Although in industrial applications these curling effects may be undesirable, they allow knits to form striking three-dimensional structures simply by varying the in-plane stitch pattern. For a plain jersey knit, yarn segments near the front of the fabric tend to be oriented in the $y$ direction while yarn segments near the back of the fabric tend to be oriented in the $x$ direction (\textbf{figure~\ref{fig:5}\textit{a}}). For a yarn with nonzero bending modulus, this difference in orientation generates a ``natural curvature'' energetically preferred by the knit stitch. At a fabric level, small knit (jersey) swatches are doubly curved, simultaneously exhibiting negative curvature in the $x$ direction and positive curvature in the $y$ direction, whereas purl (reverse jersey) swatches will have curvatures of opposite sign due to its mirror symmetry. However, across a large area of fabric, this double curvature is limited by the resistance of the fabric to stretching in the plane, a necessity dictated by Gauss's \textit{Theorema Egregium} for a defect-free planar lattice of equally sized knit stitches. Therefore, a knit fabric may generally only show double curvature near its free edges (\textbf{figure~\ref{fig:5}\textit{a}}). The bulk curvature description of knit and purl stitches supplements previous fabric-level descriptions of knitted structures using thin shell theory, enabling computational predictions of knits' ``self-folding'' patterns when these curvatures interact at the fabric scale (\textbf{figure~\ref{fig:5}\textit{b}}).

\subsubsection{Patterned geometry and mechanics}

\ejd{Beyond the local structure of stitches, larger-scale patterning of knit and purl stitches can have profound effects on the geometry and mechanics of weft-knit fabrics. These patterns can generate bulk curvature at the fabric scale, enabling extremely soft deformation modes that can be tuned for their tensile stiffness in both $x$ and $y$ directions.}
The intentional introduction of defects into the knit lattice by locally adding or removing stitches within the grid can allow for three-dimensional shape programming when applied at sufficient magnitude, with the formation of both positive and negative curvature. \cite{underwood2024design,Narayanan2018,spencer2001knitting}. There are several knitting techniques that create these defects that do \textit{not} require cutting the yarn. These include: decreases wherein a stitch is removed from a course and creates an edge dislocation (\textbf{figure~\ref{fig:5}\textit{c}}) that can result in symmetric or asymmetric stitch tilt (\textbf{figure~\ref{fig:5}\textit{d}}); increases where an additional stitch is introduced at a course (\textbf{figure~\ref{fig:5}\textit{e}}); partial or short-row knitting where a subsection of a course is knitted and adds additional \msd{stitches walewise, locally changing lattice coordination by interrupting courses and adding walewise connections}  (\textbf{figure~\ref{fig:5}\textit{g-h}}).
These defects and others are used to alter the curvature of the knitted fabric, and is one way in which socks are made to conform to the shape of our feet.
An extreme application of this is the algorithmic creation of 3D knitting patterns, such as the Stanford bunny shown in \textbf{figure~\ref{fig:5}\textit{i}}~\cite{Narayanan2018}, \msd{which is shaped through the incorporation of many edge dislocations}.
\msd{Disruptions in even spacing between courses due to defects are evocative of those seen in layered fluids (i.e.~smectic liquid crystals), which constitute a rich and ongoing field of study~\cite{Kleman2009,Severino2024}}.
The use of patterns and defects to target 3D shapes and mechanics, e.g.~for the design of bespoke garments, remains an open ``grand challenge'' in the (inverse) design of fabrics.



\section{SUMMARY AND OUTLOOK}

In this review, we have examined woven and knitted fabrics through a lens that is shaped by current interests in the geometry and topology of condensed matter as well as soft mechanical metamaterials.
Similar to the realm of polymer physics, the basic building block of a textile is not a roughly convex particle or atom, but a staple fiber or filament.
Microstructure is not directly determined by covalent or metallic interactions, but by twisted bundles, friction, links, and knots.
Continuing an ascent towards the macroscale, we described efforts to rationalize how the mechanics of yarn crossings and larger-scale patterns determines the collective mechanics and shape of the fabric.

Despite the long history of textiles, there are plenty of open questions that are amenable to the tools and concepts of condensed matter.
These questions lie at all length scales of these materials, from changes in fiber packing due to clasps between different yarn segments, to a complete characterization of defects in knitted fabrics.
There is also an opportunity to push into new frontiers of textiles by attempting to generalize the notion of a knitted stitch~\cite{takano2025chiralanaloguesknitstitches}, revisiting the design of chainmail~\cite{Klotz2024,Zhou2025}, or creating higher-dimensional generalizations of traditionally 2D textiles~\cite{Evans2013,cline2025entanglementdrivenresponsesmultiscale3dprinted}.
Finally, it is worth considering the grand questions of how to design bespoke garments and materials at the level of individual stitches and how to use these principles to design linked materials from the architectural scale to the molecular.
While there is much to explore, it is essential to remember that there is tremendous expertise that has been developed and curated within crafting circles and that much \msd{of what} we do is to simply translate that understanding into the language and formalism of our field, \msd{with the hope that we can eventually find new ways of understanding these materials.}
In doing so, there is the opportunity for physics to be reciprocally enriched by the weavers' and knitters' view of topology, symmetry, and mechanics.

\section*{DISCLOSURE STATEMENT}
The authors are not aware of any affiliations, memberships, funding, or financial holdings that might be perceived as affecting the objectivity of this review.

\section*{ACKNOWLEDGMENTS}
We thank Sabetta Matsumoto, Daria Atkinson, and James McCann for their helpful discussions and guidance.
EJD is supported by the National Science Foundation (Grant No.\ CMMI-2344589) and the College of Chemistry at UC Berkeley.
SM is supported by JSPS KAKENHI Grant-in-Aid for Early-Career Scientists (Grant Number 25K17246) and the Daiichi-Sankyo ``Habataku'' Support Program for the Next Generation of Researchers 2025.
SP acknowledges the financial support provided by JSPS KAKENHI Grant-in-Aid for Early-Career Scientists (Grant Number 25K17363). 
LN was supported by the Kaufman Foundation New Research Initiative award.  
VS acknowledges support from the National Science Foundation (Grant No.\ DMR-2502330) and the Rice University School of Engineering and Computing.


\bibliography{refs}

\end{document}